\documentclass[a4paper,twoside,10pt]{letter}
\usepackage{graphicx,saj,multicol,subeqnarray,amsmath,url}

\def\papertype{Original Scientific Paper}

\newcommand{\D}{$^\circ$}
\def\p0{\phantom{0}}
\def\it{\sl}

\def\arcmin{\hbox{$^\prime$}}
\def\arcsec{\hbox{$^{\prime\prime}$}}
\def\QSO{\mbox{{QSO~J0443.8-6141}}}

\def\udc{...}
\begin{document}
\baselineskip=3.1truemm
\columnsep=.5truecm
\newenvironment{lefteqnarray}{\arraycolsep=0pt\begin{eqnarray}}
{\end{eqnarray}\protect\aftergroup\ignorespaces}
\newenvironment{lefteqnarray*}{\arraycolsep=0pt\begin{eqnarray*}}
{\end{eqnarray*}\protect\aftergroup\ignorespaces}
\newenvironment{leftsubeqnarray}{\arraycolsep=0pt\begin{subeqnarray}}
{\end{subeqnarray}\protect\aftergroup\ignorespaces}
%


\markboth{\eightrm RADIO-CONTINUUM OBSERVATIONS OF A GIANT RADIO SOURCE \QSO}
{\eightrm M.D.~FILIPOVI{\' C}, K.O.~{\v C}AJKO, J.D. COLLIER, N.F.H. TOTHILL}

{\ }

\publ

\type

{\ }


\title{RADIO-CONTINUUM OBSERVATIONS OF A GIANT RADIO SOURCE \QSO }


\authors{M.D.~Filipovi\'c$^1$, K.O.~{\v C}ajko$^2$, J.D. Collier$^{1,3}$, N.F.H. Tothill$^1$}

\vskip 3mm


\address{$^1$University of Western Sydney, Locked Bag 1797, Penrith South DC, NSW 1797, Australia}
\Email{m.filipovic}{uws.edu.au, j.collier@uws.edu.au, n.tothill@uws.edu.au}

\address{$^2$University of Novi Sad, Faculty of Sciences, Department of Physics, 
\break Trg Dositeja Obradovi\'ca 4, 21000 Novi Sad, Serbia}
\Email{kristina.cajko}{df.uns.ac.rs}

\address{$^{3}$CSIRO Astronomy and Space Science, Marsfield, NSW 2122, Australia}


\dates{June 2013}{June 2013}


\summary{We report the discovery of a giant double-lobed (lobe-core-lobe) radio-continuum structure associated with \QSO\ at z=0.72. This QSO was originally identified during the follow-up of a sample of ROSAT All Sky Survey sources at radio and optical frequencies. With a linear size of $\sim$0.77~Mpc, \QSO\ is classified as a giant radio source (GRS); based on its physical properties, we classify \QSO\ as a FR~II radio galaxy. High-resolution observations are required to reliably identify GRSs; the next generation of southern-sky radio and optical surveys will be crucial to increasing our sample of these objects.
}


\keywords{galaxies: active --- galaxies: distances and redshifts --- galaxies: individual (\QSO)}

\begin{multicols}{2}


\section{1. INTRODUCTION}

Quasi-stellar objects (QSOs) associated with megaparsec sized giant radio sources (GRSs) are rare. GRSs are powerful extragalactic radio sources, hosted by galaxies or quasars, for which the projected linear extent (PLE) of the radio structure is $>\sim 1$~Mpc (Spergel et al. 2003) --- specifically, the PLE is $>0.72$~Mpc, using standard cosmological parameters \footnote{The cosmological constants used throughout this paper are $H_0$=67.8~kms$^{-1}$ Mpc$^{-1}$, $\Omega_{m}$=0.27 and $\Omega_{\Lambda}$=0.73.}. The largest known GRS is QSO~J1420-0545 with a PLE of 4.9~Mpc (Machalski et al.~2008). Of all known GRSs, less than 10$\%$ are QSO (Ishwara-Chandra and Saikia 1999). Ku\'zmicz \& Jamrozy (2012) recently compiled the largest sample of giant radio quasars (GRQs) to date, which included 24 new and 21 previously-known objects. They also calculated a number of important parameters of their nuclei, from which they concluded that GRQs have properties similar to quasars of smaller size. Specifically, giant quasars do not have more powerful central engines than other radio quasars. Their results are consistent with evolutionary models of extragalactic radio sources, which predict that GRSs could be more evolved (aged) sources compared to smaller radio quasars, and their large size simply reflects the longer time that the jets have been operating. Ku\'zmicz \& Jamrozy (2012) also noted that the environment may not have a significant role in the formation of large-scale radio structures. Known QSOs associated with GRSs are mostly double-lobed radio sources with weak, but detectable, core emission, and $\sim$70\% are known to be X-ray emitters. Such objects enable a number of astrophysical problems to be studied: probing of the intergalactic medium at different redshifts, developing orientation dependent unification schemes and understanding radio source evolution.

As part of a European Southern Observatory (ESO) key field program to optically identify a sub-sample of $\sim$700 ROSAT All Sky Survey (RASS) sources, Danziger et al. (1990) identified RXJ~0443.8-6141 as a QSO with redshift $z=0.72$. During the initial radio follow-up of this sub-sample at 4.8~GHz, Anderson (2002) discovered a triple (lobe-core-lobe) radio source possibly associated with RXJ~0443.8-6141. The Australia Telescope Compact Array (ATCA)\footnote{\url{http://www.narrabri.atnf.csiro.au/observing/users_guide/html/atug.html}.} data used for this initial discovery was insufficient to show if the lobes were connected to the radio core via jet structures. After encountering a similar problem with suspected triple radio source PKS~0241+011, with an estimated PLE of 2~Mpc, Nilsson \& Lehto (1997) imaged PKS~0241+011 at much higher resolution. Contrary to the hypothesis, they found that the three radio components of PKS~0241+011 were separate, physically unrelated sources. This highlights the need to thoroughly investigate the radio morphologies of suspected GRSs to unambiguously establish an association with nearby optical QSOs (Savage and Cannon 1994). 

To ensure that \QSO\ is unambiguously associated with both nearby radio lobes, observations were undertaken to map the radio morphology over a range of frequencies. This paper presents these new radio observations, as well as new optical and infrared observations of \QSO.

\section{2.	DATA}

\subsection{2.1. Radio Observations}

Radio observations of \QSO\ were taken with the ATCA at four frequencies: 1.38, 2.39, 4.8 and 8.64~GHz. These were divided into two separate ATCA projects: C133 contained observations using the array configuration 6C at frequencies of 4.8 and 8.64~GHz; C322 contained observations using configuration 6B at frequencies of 1.38 and 2.39~GHz. Both array configurations are quite sparse, and make use of a 6 km baseline, which provides high-resolution details. However, this results in the data suffering from missing short spacings, reducing the amount of extended emission that can be detected. We used the standard continuum mode with 128~MHz bandwidth, divided into 32 channels of 4~MHz each. The dual frequency ability of the ATCA allowed 1.38 and 2.39~GHz data to be recorded simultaneously, in 18 cuts evenly distributed in hour angle. The simultaneous 4.8 and 8.64~GHz data were taken over a full eight hour synthesis. Table~1 gives the observing log: Observing date, ATCA antenna configuration, observing frequency and integration time. In addition to our own observations, we made use of the Sydney University Molonglo Sky Survey (SUMSS) data (Bock et al.~1999) from the Molonglo Observatory Synthesis Telescope (MOST) at 843~MHz.


\vskip.5cm \noindent
\parbox{\columnwidth}{
{\bf Table 1.} Summary of ATCA observations used to image \QSO. \\
\vskip.5cm

\centerline{
\begin{tabular}{cccc} 
\hline 
Date of       & Array   & Freq. & Integ. Time   \\
Observation   & Config. & (GHz) & (minutes)    \\
\hline
12 Feb 1994   & 6B      & 1.38  & \p055         \\
12 Feb 1994   & 6B      & 2.39  & \p055        \\
\p04 Sep 1993 & 6C      & 4.80  & 485           \\
\p04 Sep 1993 & 6C      & 8.64  & 485           \\
\hline
\end{tabular}}} \vskip.5cm


\subsection{2.2. Data Reduction}

The \textsc{miriad} (Sault et al.~1995) and \textsc{karma} (Gooch~1995) software packages were used for reduction and analysis of the radio-continuum data. More information on the observing procedure can be found in Boji\v{c}i\'c~et~al.~(2007), Crawford~et~al.~(2008a,b; 2010) and \v{C}ajko~et~al.~(2009). Two standard sources were used as calibrators for creating images: Primary calibrator 1934-638 was used for flux and bandpass calibration and secondary calibrator \mbox{0522-611} was used for phase calibration. 
					
To create the image at 1.38~GHz, all data from the two observing sessions for project C322 (12$^{\mathrm {th}}$ and 16$^{\mathrm {th}}$ February~1994) were merged in the {\it uv} plane. The dirty map was produced using multi-frequency synthesis imaging (Sault and Wieringa 1994) with natural weighting. To deconvolve the dirty map, the \textsc{miriad} task {\tt mossdi} was used. This algorithm is suitable for deconvolving mosaics which consist of point source emissions detected with long baselines. The dirty map was \textsc{clean}ed until the side lobes of strong point sources were no longer present. The restored image has a synthesised beam of 11.5\arcsec\ and r.m.s.\ noise of 0.7~mJy/beam. The restored image at 2.39~GHz was constructed using the same procedure, and has a synthesised beam of 7.5\arcsec\ and r.m.s.\ noise of 0.3~mJy/beam. 
					
The images at 4.8~GHz and 8.64~GHz were produced by the same procedure, using observations combined from the two observing sessions for project C133 (26$^{\mathrm{th}}$ March and 4$^{\mathrm{th}}$ September 1993). The restored 4.8~GHz image has a synthesised beam size of 3.1$\times$1.7\arcsec, a PA of --4.2\D\ and r.m.s.\ noise of 0.04~mJy/beam. The restored 8.64~GHz image has a synthesised beam size of 1.93$\times$1.09\arcsec, a PA of --3.9\D\ and r.m.s.\ noise of 0.04~mJy/beam. All images were corrected for the attenuation of the primary beam using the task {\tt linmos}.

Additional maps were created at each frequency by re-gridding the images with the task {\tt regrid}, to match the size and resolution of the 1.38~GHz image, which had the poorest resolution, thus allowing no oversampling to occur. A spectral index map was then created using these re-gridded maps from all four observed frequencies. This was done using the \textsc{miriad} task {\tt maths}, which calculated the spectral index\footnote{spectral index $\alpha$ is defined by $S_{\nu}=\nu^{\alpha}$, where $S_{\nu}$ is the integrated flux density and $\nu$ is the frequency.}  ($\alpha$) of each pixel above a level of 5$\sigma$ (0.1 -- 1~mJy/beam). Pixels below this level were blanked in the spectral index map.

\subsection{2.3. Infrared and Optical Observations}

Infrared images centred on \QSO\ were extracted from the datasets of the Wide-Field Infrared Survey Explorer (WISE; Wright et al. 2010). WISE obtained all-sky surveys centred at 3.4, 4.6, 12 and $22\,\mu$m, with respective angular resolutions of 6.1$\arcsec$, 6.4$\arcsec$, 6.5$\arcsec$ and 12$\arcsec$, reaching respective average $5\sigma$ sensitivity levels of 0.08, 0.11, 1 and $6\,$ mJy, with sensitivity increasing towards the ecliptic poles. 

Optical images were extracted from the ESO Online Digitized Sky Survey\footnote{\url{http://archive.eso.org/dss/dss}}. We make use of an image from the UK Schmidt Telescope's blue band.

\section{3. RESULTS AND ANALYSIS}

In Figs. 1, 2, 3 and 4 we show radio-continuum images of \QSO\ created at 1.38, 2.39 and 4.8~GHz, along with the WISE and DSS-blue images. The spectral index map created between 1.38 and 4.8~GHz is shown in Fig.~5. From these images, it can be seen that \QSO\ consists of three radio structures, which appear to be the central AGN core and two symmetrically positioned lobes of a radio galaxy (see Table~2). In this scenario, the southern jet (L2; Table~2) seems stronger than the northern one (L1; Table~2), causing its lobe to be located much closer to the core. Some 4\arcsec\ to the east (from the central source; marked as BG2 in Table~2), we identify another point-like source as a non-related background object with a steep spectral index of \mbox{$\alpha$=--0.88$\pm$0.36}. We also note its co-identification at optical wavelengths, seen in the DSS blue band in Fig.~3.

The greyscale in Fig.~1 shows \QSO\ as it appears in the SUMSS image, at a resolution of $\sim$45$\arcsec$. The higher resolution 1.38~GHz image shown by the contours reveals that there is a morphological connection between these two components seen in SUMSS. Fig.~2 shows the source at 4.8~GHz with 2.39~GHz contours overlaid. The optical and infrared observations of \QSO\ from DSS and WISE are shown in Figs.~3 and 4, respectively overlaid with 4.8 and 2.39~GHz contours. These images reveal that the central radio source is coincident with an optical and infrared source. This shows that the central radio source is the AGN core of \QSO, located within the host galaxy detected in the optical and infrared. The absence of any optical and infrared emission from the extended radio emission towards the north and south of the AGN core is consistent with the radio emission originating from two radio lobes powered by two symmetrically-positioned jets, since no significant thermal emission is expected from the lobes of radio galaxies. Furthermore, as seen from the 4.8~GHz contours in Fig.~3, two compact regions of radio emission can be found within the lobes. These compact regions are typical of the hotspots seen in Fanaroff-Riley Type~II (FR~II; Fanaroff \& Riley 1974) radio galaxies. After fitting a spectral index of $\alpha=-0.9$ to the entire radio galaxy, we extrapolated a 178~MHz flux density of $\sim$0.65~Jy. At a redshift of $z=0.71$, this corresponds to a luminosity of $P_{\mbox{\small{178~MHz}}} \sim 1.7 \times 10^{27}$~W\,Hz$^{-1}$, well above the FR~I/FR~II break luminosity of  2$\times 10^{25}$ W\,Hz$^{-1}$. Hence, we conclude that \QSO\ is a FR~II radio galaxy.

At 4.8~GHz, using a line profile across the two radio lobes, we measure the separation between the two most distant points in the lobes detected above 5$\sigma$ as $\sim$102$\arcsec$. At a redshift of $z=0.72$, the cosmological parameters used in this paper give a scale of 7.6 kpc/$\arcsec$. Therefore, \QSO\ has a PLE of 0.77~Mpc, just large enough to be classified as a GRS.

We divide \QSO\ into the three regions listed in Table~2: the northern region listed as L1 (lobe one), the central core, and southern region listed as L2 (lobe two). We also list the two nearby unrelated sources in Table~2: the strong compact source located to the East of the core of \QSO\, listed as BG1, and the weak source located even closer to the East of the core, detected only at 4.8~GHz, labelled as BG2. We measured flux densities from each of our radio-frequencies as well as spectral indices for each component/region presented in Table~2. All flux density measurements have an associated error of $<$10\%. However, we point that the error in flux density is intensity dependent where stronger sources are associated with smaller uncertainties.

\end{multicols}

\centerline{\includegraphics[angle=-90,width=1.6\textwidth]{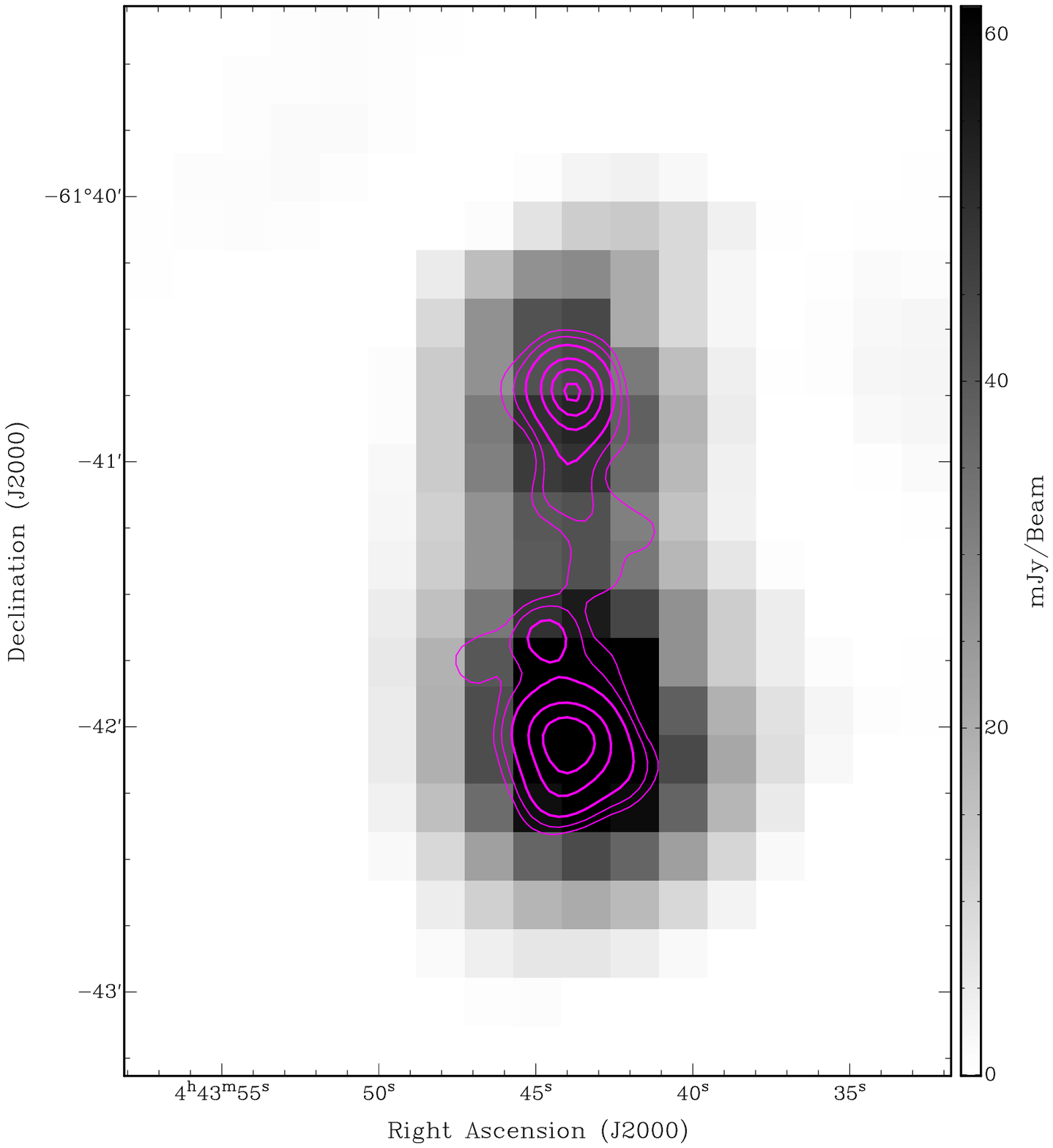}}
\label{SUMMS}
\figurecaption{1.}{SUMMS 843~MHz ($\lambda$=36~cm) image of \QSO\ overlaid with ATCA contours at 1.38~GHz ($\lambda$=20~cm). The side bar shows the greyscale in units of mJy/beam. The 843~MHz and 1.38~GHz r.m.s.\ noises are respectively 2 and 0.7~mJy/beam. The 1.38~GHz contours levels are 2, 3, 5, 10, 15 and 20~mJy/beam. 
}

\centerline{\includegraphics[angle=-90,width=1.6\textwidth]{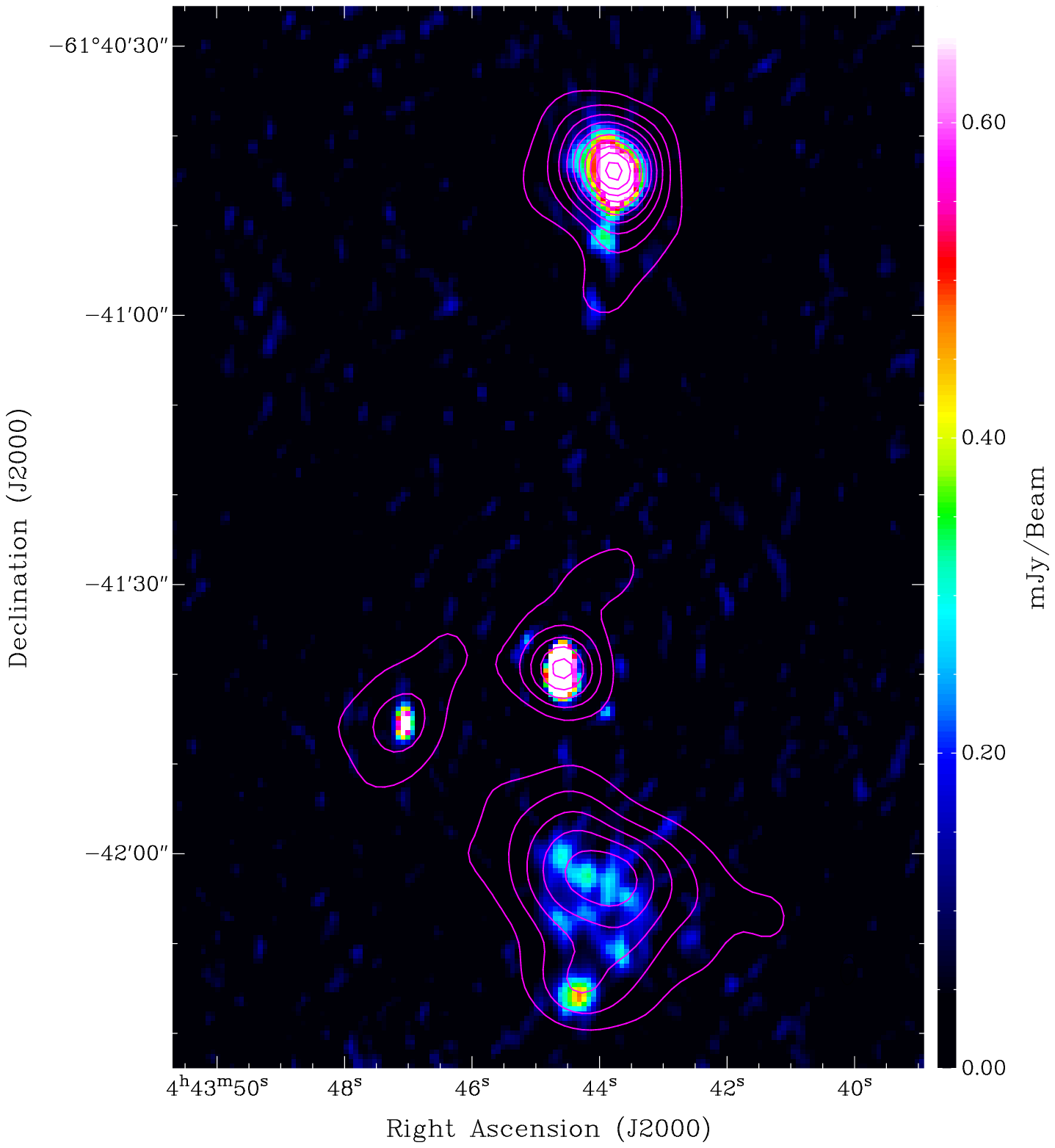}}
\label{6cm}
\figurecaption{2.}{ATCA observations of \QSO\ at 4.8~GHz ($\lambda$=6~cm) overlaid with 2.39~GHz ($\lambda$=13~cm) contours. The sidebar shows the colour scale in units of mJy/beam. The 4.8 and 2.39~GHz r.m.s.\ noises are respectively 0.04 and 0.3~mJy/beam. The 2.39~GHz contour levels are 0.9, 1.8, 2.7, 3.6, 4.5, 5.4, 6.3, 7.2, 8.1 and 9.0~mJy/beam. 
}

\centerline{\includegraphics[angle=-90,width=1.6\textwidth]{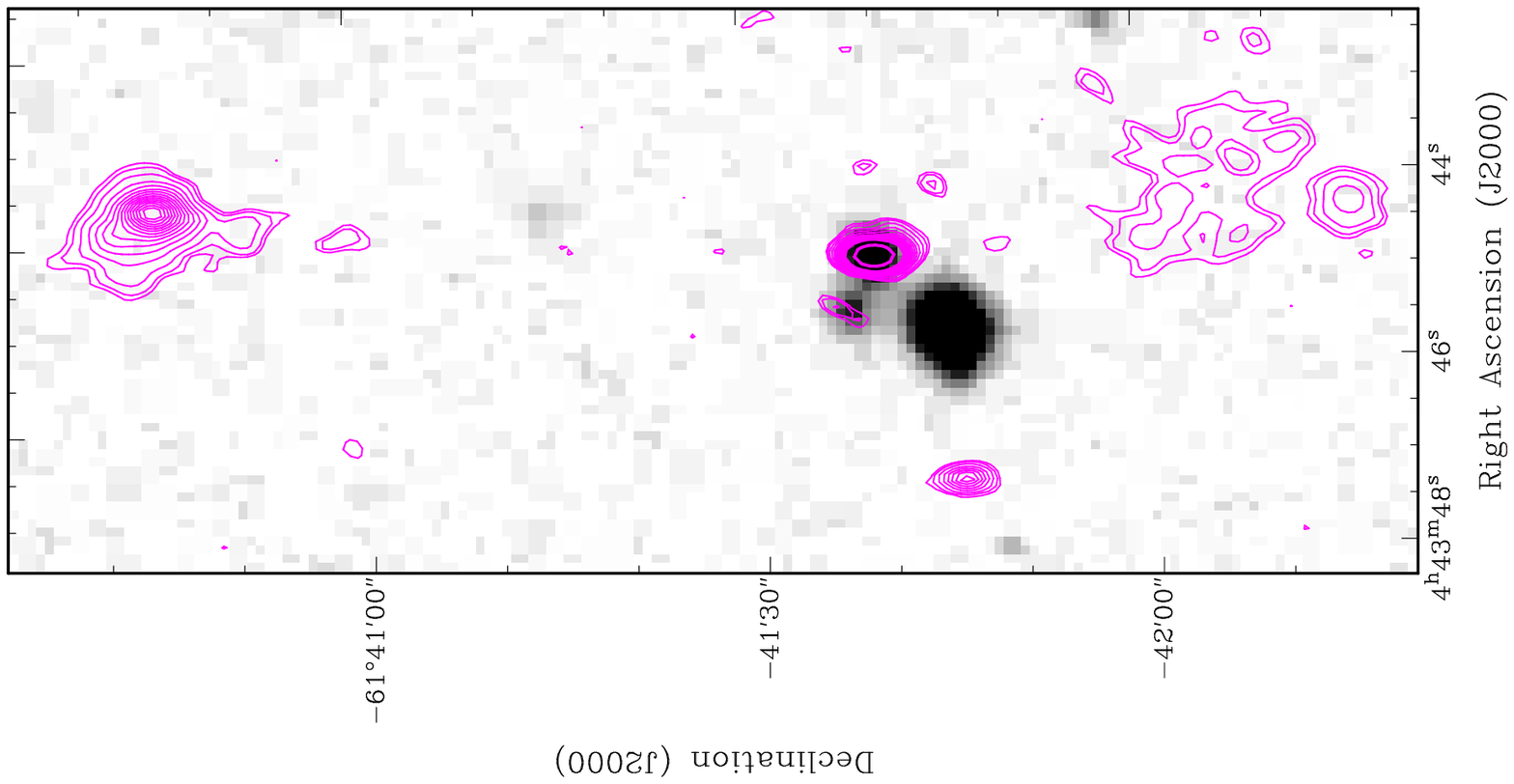}}
\label{DSS}
\figurecaption{3.}{The DSS2-blue image (gray scale) of \QSO\ overlaid with 4.8~GHz ($\lambda$=6~cm) contours. The 4.8~GHz contours levels are from 0.12 to 5~mJy/beam, in steps of 3$\sigma$ (0.12~mJy/beam). 
}

\centerline{\includegraphics[angle=-90,width=1.5\textwidth]{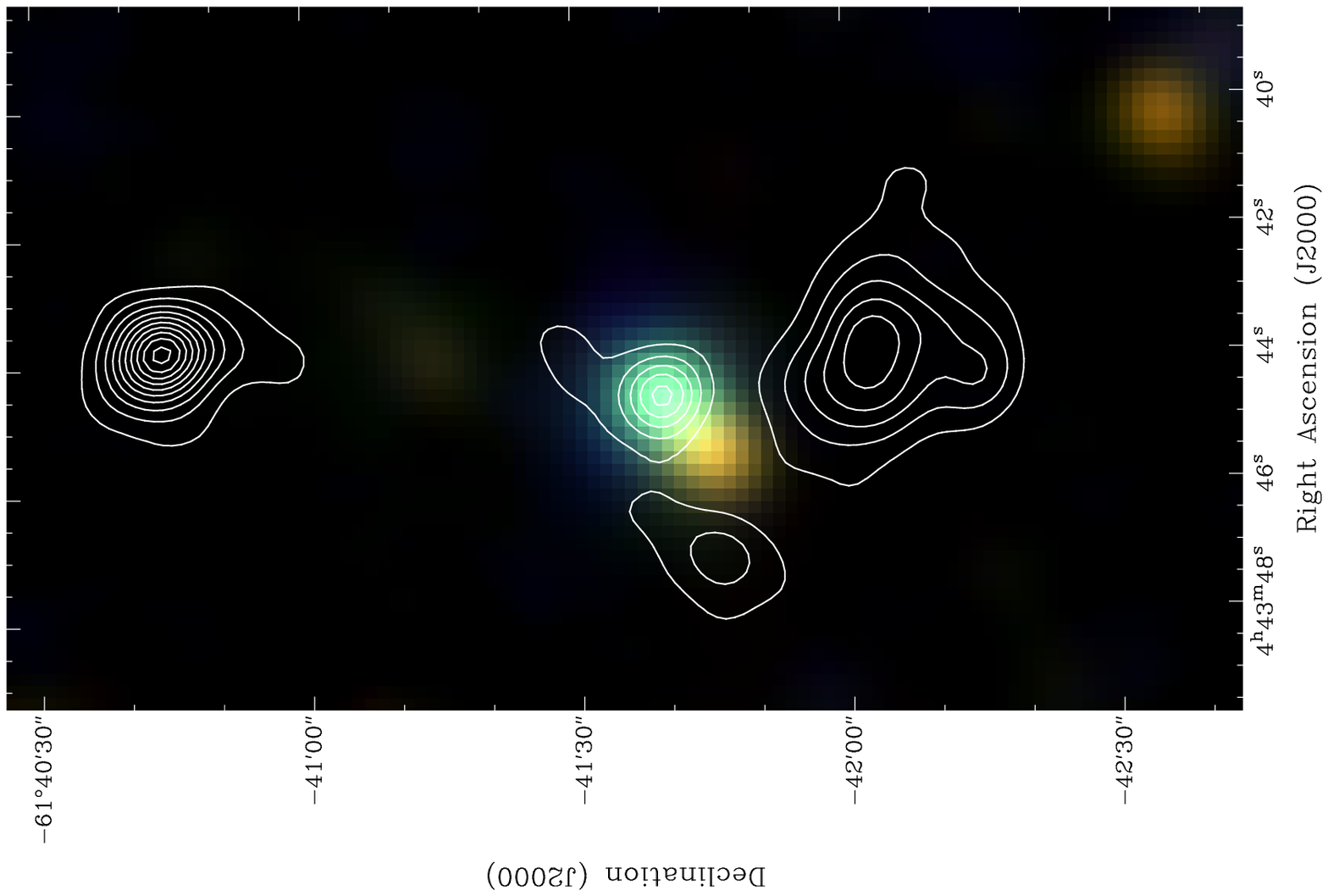}}
\label{WISE}
\figurecaption{4.}{WISE colour-composite image of \QSO\ overlaid with 2.39~GHz ($\lambda$=13~cm) contours. The 2.39~GHz contour levels are 0.9, 1.8, 2.7, 3.6, 4.5, 5.4, 6.3, 7.2, 8.1 and 9.0~mJy/beam. 
The WISE image uses the 3.4, 4.6 and 12 $\mu$m bands for blue, greed and red channels respectively.
}


\vskip.5cm \noindent
\parbox{\textwidth}{
{\bf Table 2.} Measured flux densities and spectral indices for three components of \QSO\ and two unrelated nearby sources. All flux density measurements have an associated error of $<$10\%. The error in spectral index represent a statistical uncertainty based on upper flux density error of $<$10\%.\\

\centerline{
\begin{tabular}{lccccccccccc} 
\hline 
Object & RA      & Dec      & $S_{0.843}$ & $S_{1.38}$ & $S_{2.39}$ & $S_{4.8}$ & $S_{8.64}$  & $\alpha \pm \Delta \alpha$ \\
       & (J2000) & (J2000)  & (Jy)       & (Jy)       & (Jy)       & (Jy)      &  (Jy)       &    \\
\hline 
Core	& 4$^{h}$43$^{m}$44.6$^{s}$ & --61\D41\arcmin39.32\arcsec & 0.006 & 0.0063 & 0.0046 & 0.0078\p0 & 0.0050 & $+0.0\pm$0.2 \\
L1  & 4$^{h}$43$^{m}$43.7$^{s}$ & --61\D40\arcmin44.45\arcsec & 0.063 & 0.0320 & 0.0150 & 0.0083\p0 & 0.0027 & $-1.3\pm$0.2 \\
L2  & 4$^{h}$43$^{m}$43.8$^{s}$ & --61\D42\arcmin02.91\arcsec & 0.091 & 0.0550 & 0.0220 & 0.0074\p0 & 0.0030 & $-1.5\pm$0.2 \\
BG1 & 4$^{h}$43$^{m}$47.1$^{s}$ & --61\D41\arcmin45.40\arcsec & ---   & 0.0026 & 0.0024 & 0.0009\p0 &  ---   & $-0.9\pm$0.4 \\
BG2 & 4$^{h}$43$^{m}$45.1$^{s}$ & --61\D41\arcmin36.36\arcsec & ---   & ---    &   ---  & 0.00025   &  ---   & --- \\
\hline
\end{tabular}}} \vskip.5cm

\begin{multicols}{2}

The spectral index data (Table~2 and Fig.~5) show that the core of \QSO\ has a flat spectrum ($\alpha\sim$ 0), while the northern and southern lobes have much steeper spectral indices. The flat spectral index from the central source is what is expected from an AGN core, since the core is dominated by the thermal emission of the accretion disc. The steep spectral indices of the lobes are likewise expected since these regions mark the place in which the relativistic jets collide with the intergalactic medium, causing strong non-thermal synchrotron emission. The southern lobe L2, located closer to the core of \QSO, has a slightly steeper spectral index of $\alpha=-1.5$, compared to the northern lobe L1, which is located somewhat further from the core and has a spectral index of $\alpha=-1.3$. This is consistent with the hypothesis from Ku\'zmicz \& Jamrozy (2012), suggesting that the environment of GRQs plays little role in the morphology of the radio galaxy --- L2 might be expected to have a denser environment, and hence a flatter spectral index, but this is not observed. \QSO\ may be a radio galaxy with a complex morphology (such as a Wide Angle Tail) for which orientation effects cause different geometrical projections to be observed.

\end{multicols}
\centerline{\includegraphics[width=0.75\textwidth]{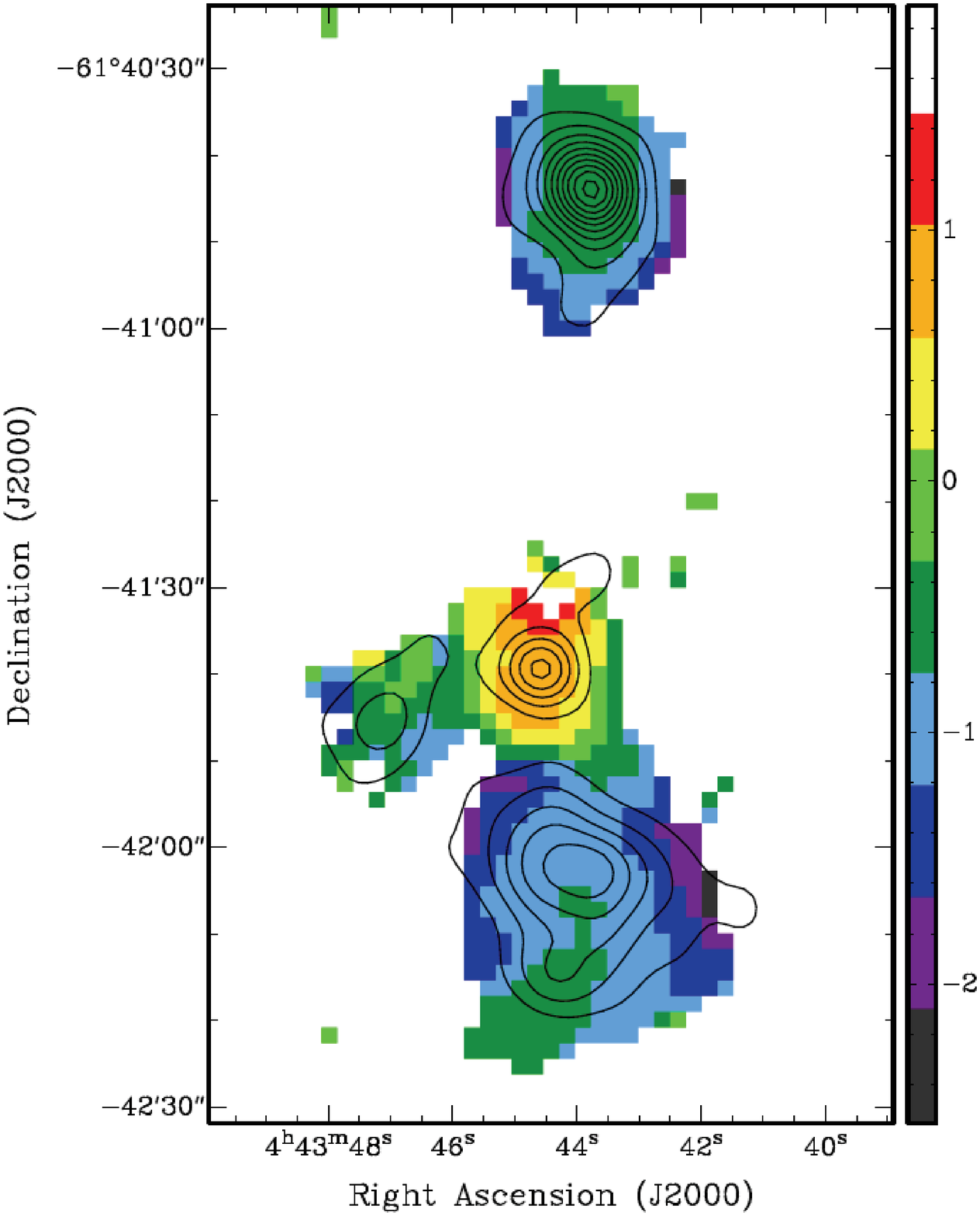}}
\figurecaption{5.}{Spectral index map of \QSO\ calculated between 1.38 and 4.8~GHz, overlaid with 2.39~GHz contours. The colour bar quantifies the spectral index at each pixel above 5$\sigma$ level ($0.1 -- 1$~mJy/beam). The contour levels are the same as in Fig.~2.}
\begin{multicols}{2}

\section{4. CONCLUSION}

We confirm the existence of a GRS associated with \QSO, which is located at a redshift of $z=0.72$ and has a PLE of 0.77~Mpc. We conclude from the radio morphology and radio luminosity that \QSO\ is an FR~II radio galaxy. This identification is only made possible by using radio observations taken at various frequencies of different resolution. While the lower frequency data enabled accurate measurements of the total integrated flux, and revealed the physical connection between the AGN core and lobes, the higher frequency data enabled the compact hot spots typical of FR~II galaxies to be identified. Using optical and infrared images also enabled unrelated background sources to be identified. Spectral index measurements taken between all radio frequencies were able to confirm these as unrelated background sources, and confirm our hypothesis that \QSO\ is a lobe-core-lobe GRS. Compared to the rest of the GRQ sample (Ku\'zmicz \& Jamrozy 2012), \QSO\ has a fairly average redshift, and is at the small end of the scale in PLE. 

Ku\'zmicz \& Jamrozy (2012) approximately doubled the number of known GRQs using the main northern hemisphere optical and radio surveys: SDSS and FIRST. All of their new sources have radio data from FIRST (rather than the less sensitive NVSS), and the new sources are generally at higher redshift. This demonstrates the importance of high-sensitivity, high-resolution (few arcsec) hemisphere-scale surveys in finding GRQs, in particular the more distant ones. Such surveys will become available in the southern hemisphere with the commissioning of instruments such as ASKAP, which will deliver few-arcsecond resolution and high sensitivity over the entire southern sky. These surveys will be the next major step in finding GRQs, although follow-up observations over multiple frequencies may be required to obtain the spectral index maps show lobe-core-lobe structure.

\vskip-1mm


\acknowledgements{We used the {\sc karma} and {\sc miriad} software package developed by the ATNF. The Australia Telescope Compact Array is part of the Australia Telescope which is funded by the Commonwealth of Australia for operation as a National Facility managed by CSIRO. We thanks Martin W. B. Anderson for the initiation of this project. 
}

\vskip.6cm


\references

\vskip-1mm

Anderson, M. W. B.: 2002, in A Radio Survey Of Selected Fields From The ROSAT All Sky Survey, Ph. D. Thesis, University of Western Sydney.

Bock, D. C.-J., Large, M. I., Sadler, E. M.: 1999, \journal{Astron. J.}, \vol{117}, 1578.

Boji\v{c}i\'c, I. S., Filipovi\'{c}, M. D., Parker, Q. A., Payne, J. L., Jones, P. A., Reid, W., Kawamura, A. and Fukui, Y.: 2007, \journal{Mon. Not. R. Astron. Soc.}, \vol{378}, 1237.

Crawford, E. J., Filipovi\'{c}, M. D. and Payne, J. L.: 2008a, \journal{Serb. Astron. J.}, \vol{176}, 59. 

Crawford, E. J., Filipovi\'{c}, M. D., De Horta, A. Y., Stootman, F. H. and Payne J. L.: 2008b, \journal{Serb. Astron. J.}, \vol{177}, 61.

Crawford, E. J., Filipovi\'{c}, M. D., Haberl, F., Pietsch, W., Payne, J. L. and De Horta, A. Y.: 2010, \journal{Astron. Astrophys.}, \vol{518}, A35.

\v{C}ajko, K. O., Crawford, E. J., Filipovi\'{c}, M. D.: 2009, \journal{Serb. Astron. J.}, \vol{179}, 55.

Danziger, I. J., Tr$\ddot{u}$mper, J., Beuermann, K., B$\ddot{o}$hringer, H., Flemming, T., Gottwald, M., Hasinger, G., Krautter, J., Macgillivray, H., Miller, L., Pakull, M., Parker, Q., Pasquini, L., Reinsch, K., Thomas, H. C., Ulrich M. H., Voges, W. and Zimmermann, H. U.: 1990, \journal{ESO Messenger}, \vol{62}, 4.

Fanaroff, B. L., Riley, J. M.: 1974, \journal{Mon. Not. R. Astron. Soc.}, \vol{167}, 31.

Gooch, R. E.: 1995, in Astronomical Data Analysis Software and Systems IV, ASP Conf. Series ed. Shaw, R. A., Payne, H. E. and Hayes, J. J. E., \vol{77}, 144. 

Ishwara-Chandra, C. H. and Saikia, D. J.: 1999, \journal{Mon. Not. R. Astron. Soc.}, \vol{309}, 100.

Ku\'zmicz, A. and Jamrozy, M.: 2012, \journal{Mon. Not. R. Astron. Soc.}, \vol{426}, 851.

Machalski, J., Koziel-Wierzbowska, D., Jamrozy, M. and Saikia, D. J.: 2008, \journal{Astrophys. J.}, \vol{679}, 149.

Nilsson, K. and Lehto, H. J.: 1997, \journal{Astron. Astrophys.}, \vol{328}, 526.

Sault, R. J., Teuben, P. J. and Wright, M. C. H.: 1995, in Astronomical Data Analysis Software and Systems IV ASP Conference Series, ed. Shaw, R. A., Payne, H. E. and Hayes, J. J. E., \vol{77}, 433.

Sault, R. J. and Wieringa, M. H.: 1994, \journal{Astron. Astrophys. Suppl. Ser.}, \vol{108}, 585.

Savage, A. and Cannon, R. D.: 1994, in: The future utilisation of Schmidt telescopes. Astronomical Society of the Pacific Conference Series, Volume 84; Proceedings of IAU Colloquium 148 held 7-11 March 1994 in Bandung (Indonesia), edited by J. Chapman, R. Cannon, S. Harrison and B. Hidayat, p. 245.

Spergel, D. N., Verde, L., Peiris, H. V., Komatsu, E., Nolta, M. R., Bennett, C. L. , Halpern, M., Hinshaw, G., Jarosik, N., Kogut, A., Limon, M., Meyer, S. S., Page, L., Tucker, G. S., Weiland, J. L., Wollack, E. and Wright, E. L.: 2003, \journal{Astrophys. J. Suppl. Ser.}, \vol{148}, 175.

Wright, E. L., et al.: 2010, \journal{Astron. J.}, \vol{140}, 1868.

\endreferences 
\end{multicols}

\vfill\eject

{\ }



\naslov{RADIO-KONTINUM POSMATRANJA VELIKOG RADIO IZVORA --- \textrm{\QSO}}


\authors{M.D.~Filipovi\'c$^1$, K.O.~{\v C}ajko$^2$, J.D. Collier$^{1,3}$, N.F.H. Tothill$^1$}
\vskip3mm


\address{$^1$University of Western Sydney, Locked Bag 1797, Penrith South DC, NSW 1797, Australia}
\Email{m.filipovic}{uws.edu.au j.collier@uws.edu.au n.tothill@uws.edu.au}

\address{$^2$University of Novi Sad, Faculty of Sciences, Department of Physics, 
\break Trg Dositeja Obradovi\'ca 4, 21000 Novi Sad, Serbia}
\Email{kristina.cajko}{df.uns.ac.rs}

\address{$^{3}$CSIRO Astronomy and Space Science, Marsfield, NSW 2122, Australia}

\vskip.7cm


\centerline{UDK \udc}



\vskip.7cm

\begin{multicols}{2}
{


{\rrm U ovoj studiji predstav{lj}amo otkri{\cc}e ogromne dvostruke radio strukture koja je povezana sa kvazarom \textrm{\QSO} qiji je crveni pomak \textrm{z=0.72}. Ovaj objekat je identifikovan tokom radio i optiqkih posmatranja poznatih  \textrm{ROSAT~All~Sky~Survey} izvora. Sa veliqinom od 0.77~\textrm{Mpc} je \textrm{FR~II} tip gigant radio galaksije (GRS) a ujedno se nalazi me{dj}u uda{lj}enijim poznatim objektima ove vrste. Posmatra{nj}a u visokoj rezoluciji su neophodna za detekciju ovih objekata. Oqekujemo da {\cc}e pregledi ju{\zz}nog neba slede{\cc}e generacije identifikovati veliki broj ovakvih objekata.}

}
\end{multicols}
\end{document}